# Influence of noncontact dissipation in the tapping mode: Attempt to extract quantitative information on the surface properties with the local force probe method

J. P. Aimé,[a)] R. Boisgard, L. Nony, and G. Couturier
*Université Bordeaux I, 351 Cours de la Liberation, F-33405 Talence, France*

In the Tapping mode, a variation of the oscillation amplitude and phase as a function of the tip sample distance is the necessary measurement to access quantitatively to the properties of the surface. In the present work, we give a systematic comparison between experimental data recorded on two surfaces, phase and amplitude, and theoretical curves. With an interaction between the tip and the surface taking into account an attractive and a repulsive term, the analytical approach is unable to properly describe the relationship between the phase variation and the oscillation amplitude variation. When an additional dissipation term is involved, due to the attractive interaction between the tip and the surface, the model gives a good agreement with the recorded data. Particularly, the trends in the phase variations related to the noncontact situations have been found to be amenable to an analysis based upon a simple viscoelastic behavior of the surface. []

## I. INTRODUCTION

There are numerous experimental evidence that dynamic force microscopy is an appropriate tool to probe nanomechanical properties of soft objects at the nm scale. Experimentally, the use of an oscillating tip-cantilever system (OTCL) to probe surface properties at the local scale, from the nanometer to the picometer, is done with two different operating modes.

One mode keeps the oscillating amplitude constant (NC-AFM), and recording image is obtained by moving up and down the surface to keep a chosen resonance frequency shift constant. The experiment is performed without any contact between the tip and the surface.[1–5] With the second mode, a drive frequency is chosen. The feedback loop is used to maintain constant the amplitude of the OTCL. The recorded images are the vertical displacements needed to keep the oscillation amplitude constant. This mode, commonly called Tapping, is often used in intermittent contact (IC), that is, during a part of the oscillating period the tip touches the surface but images can also be recorded without any contact. This mode had been conceived mainly to reduce the shear forces at the interface between the tip and the surface. Companion theoretical developments demonstrate that the high sensitivity of these two modes is due to the nonlinear dynamical behavior of the OTCL at proximity of the surface.[6–9]

Therefore, a new area was open in which soft materials, polymers, and biological systems can be investigated without producing significant damages. Numerous experimental results have shown the ability of this mode to image soft materials. Among them, images of copolymers are quite convincing of its great potentiality.[10–13] However, recording a true topography is far to be achieved, it is often worth discussing image as a function of nanomechanical properties of the sample probed by this mode. Several theoretical approaches have been dedicated to the Tapping,[14–19] some of them being numerical simulations. For example, phase contrast can be explained in terms of energy dissipation into the tip–sample contact.[20,21]

The main difference between the two modes is a purely technical one and only concerns the different ways changes of the oscillating behavior as a function of the tip surface distance are detected. The Tapping mode records amplitude and phase variations while the NC AFM records resonance frequency shift and damping coefficient variations.

Besides, when the tip approaches the surface, the attractive force between the tip and the sample can be as high as 1 nN, a rather large force. Therefore, one has to take into account the work performed on the surface and a possible dissipation even without any contacts between the tip and the surface.

In NC-AFM, experiments show a change of the damping coefficient that depends abruptly on the tip–sample distance.[4,22–26] Since the tip does not touch the sample, a question rises on the physical origin of the increase of the loss of energy. A few recent works have been specially dedicated to the study of the microlever energy loss in NC-AFM.[22–26] In Ref. 26, the local deformation of the sample under the action of the oscillating tip is considered as being the leading term to explain the physical origin of the additional dissipation. A comparison between the NC-AFM results performed on a graphite surface and theoretical predictions provide an excellent agreement.[26]

The present paper is an attempt to derive analytical expressions for the Tapping mode describing the influence of the mechanical properties of the sample as an additional dissipation term in noncontact situations. There are several reasons that make an analytical description of the local sample

[a)]Electronic mail: jpaime@frbdx11.cribx1.u-bordeaux.fr

properties difficult to achieve. The first one is an appropriate description of the locality of the mechanical response of the surface; in many cases numerical simulations are required based on *ab initio* calculations.[27,28] The second is a proper description of the action of the oscillating nanotip above the sample. Let us consider that the tip sample interaction is correctly described with a power law $C/[D-x(t)]^n$, where $C$ and $n$ are functions of the type of interaction and of the geometry and size of the tip. $D$ is the distance between the cantilever at rest and the surface and $x(t)$ is the tip location ($x(t)$ can be suitably described with $x(t)=A(D)\cos[\omega t+\phi(D)]$[17,19]). An exact description of the force acting on the sample requires Fourier series leading to a rather complex mathematical development. Nevertheless, by considering the asymptotic regimes analytical expressions are obtained, allowing the experimental results to be fitted and, in turn, providing the opportunity to extract quantitative information from AFM measurements. The characteristic time scale of the sample relaxation controls the two asymptotic regimes. With a sample relaxation time much greater than that of the oscillation period (a few microseconds), the action of the oscillating tip can be reduced to the zero frequency component of the Fourier series. Such an approximation is suitable for highly viscous materials like a glassy polymer. In Ref. 29, a simple approximation was employed that describes the action of the oscillating tip as a rectangular periodic function, then the variation of the oscillation amplitude is interpreted as the result of the growth of a polymer nanoprotuberance under the action of the oscillating tip. The characteristic time of the polymer being larger than that of the oscillation period of the cantilever, the action of the tip was reduced to its average static component. Such an approximation allows an analytic expression to be derived from a self-consistent equation that describes the viscoelastic behavior of the polymer nanoprotuberance. The opposite situation occurs when relaxation times are faster than that of the oscillation period, as it happens with a graphite surface. In that case it is easy to show that the action of the tip can be suitably ascribed as a pulse.[26] Then using the Fourier transform of a pulse, an analytical expression is obtained that explains the additional dissipation as a direct consequence of the local mechanical response of the surface.[26] This latter approach is used as an attempt to describe the influence of the NC dissipation in Tapping mode measurements on hard surfaces.

Our first goal in the present work is an attempt to understand the origin of the discrepancy between theoretical development based on the Lagrangian formalism and the experimental data. While the general nonlinear behavior of the oscillator at the proximity of the surface is properly described,[7,17,19] there were still robust quantitative discrepancies between predicted variations and experimental ones. For example, theoretical curves always predicted a hysteresis loop in which the amplitude must reach the resonance one, which is never observed experimentally.[17] Also, the relationship between the amplitude and phase jumps at the bifurcation spot cannot be understood by uniquely considering the attractive and repulsive interaction in intermittent contact situations. Such discrepancies clearly show that a physical process was not considered in the previous analysis.[19]

In the present paper we first discuss the level of approximation required to use the Lagrangian formalism, and the usefulness of such an approximation for experimental results. Then a comparison with recorded approach–retract curves giving the variation of the amplitude and phase as a function of the cantilever sample distance is done (Fig. 1). Such curves, similar to the force curves in contact AFM, are necessary preliminary experiments to choose the experimental conditions to record images. A special discussion will be dedicated to noncontact and intermittent contact situations. When intermittent contact situations occur, analytical expressions can be derived both for the phase and the oscillation amplitude, while for noncontact situations an analytical expression is only obtained for the variation of the phase.

The last part of the paper is dedicated to a comparison between the theoretical development and experimental results. Two surfaces were investigated: a silica surface and a grafted surface with Aminopropylsilanes (APTES). The analysis of the experimental data is followed by an attempt to extract quantitative information.

## II. MODELING THE OTCL'S BEHAVIOR

The present paragraph is dedicated to a description of the approximation used, allowing analytical expressions to be derived to fit the experimental data. Theoretical curves are calculated for driven frequency slightly below the resonance one, but the equations can be used for any chosen driven frequency and also for the NC resonant contact mode.[7,26] The computation at a driven frequency below the resonance one provides, first, an easy way to determine the experimental conditions separating the noncontact and intermittent contact situations.[19] Second, from a practical point of view the use of a driven frequency slightly below the resonance one is the most accurate way to locate the surface (see Sec. III).

### A. Attractive regime

In this part we recall the results describing variations of the phase and of the oscillation amplitude in the dominant attractive regime.[19] Using a sphere-plan geometry with an attractive van der Waals interaction, the attractive force between the tip and the surface is

$$F_{\text{Attractive}}[x(t)] = -\frac{HR}{6[D-x(t)]^2}, \tag{1}$$

where $H$ is the Hamaker constant, $R$ the tip's apex radius, $D$ the distance between the sample and the equilibrium position at rest of the OTCL, and $x(t)=A\cos(\omega t+\phi)$, the location of the tip at time $t$. The principle of least action leads to the two equations:[19]

$$\cos\phi = Qa(1-u^2) - \frac{Q\kappa_a}{3}\frac{a}{(d^2-a^2)^{3/2}}, \tag{2a}$$

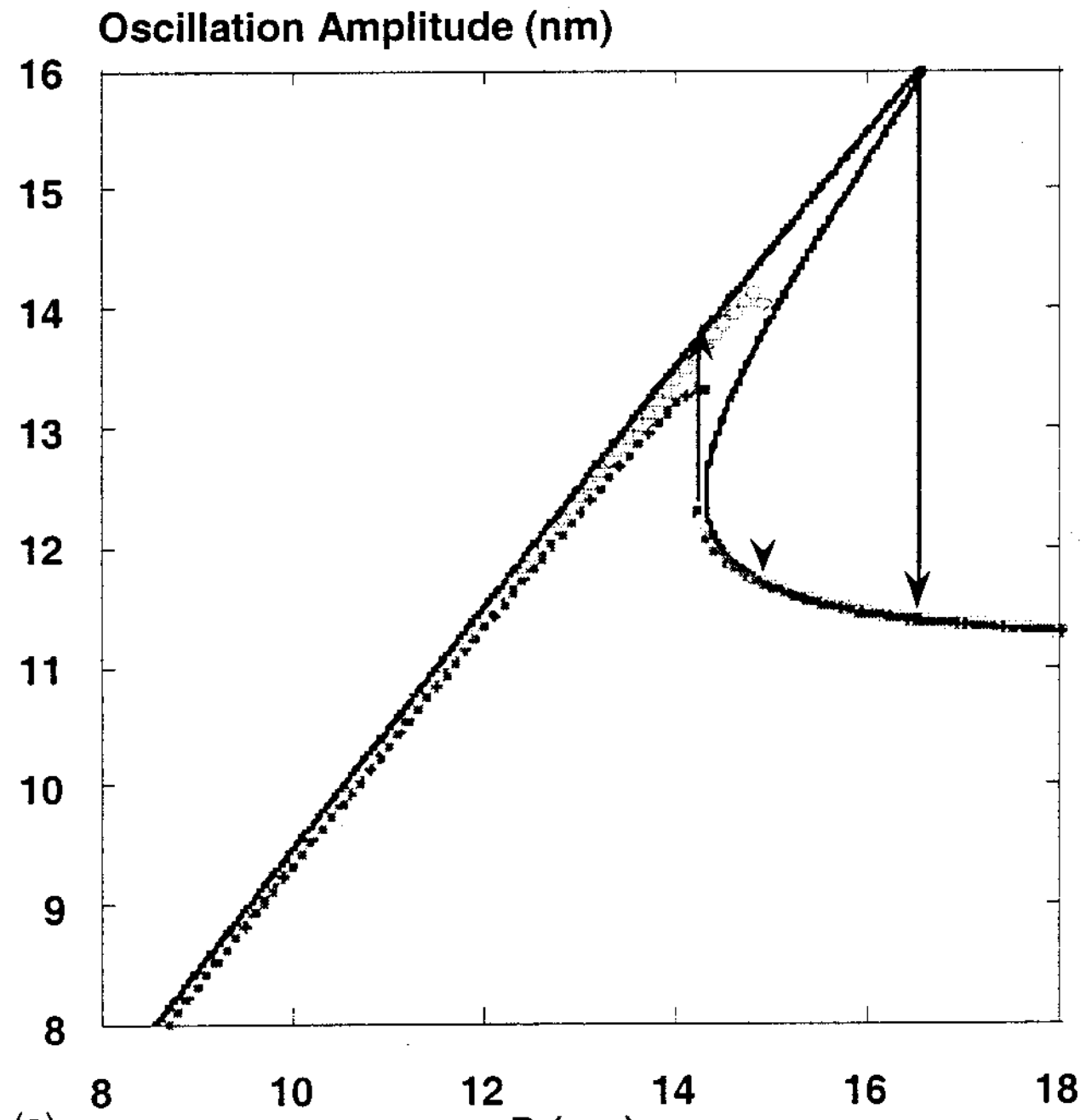

(a)

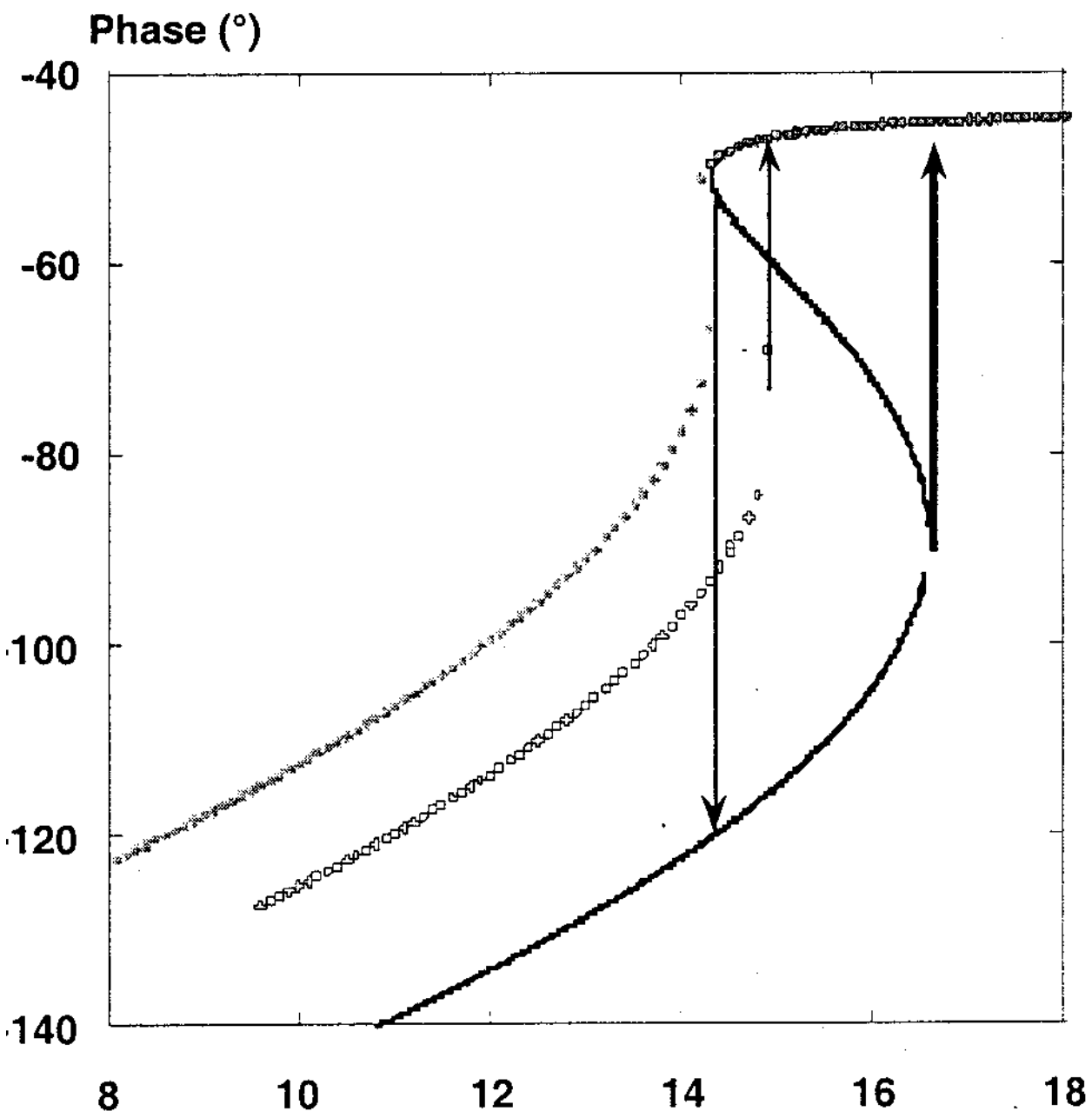

(b)

FIG. 1. (a) Calculated approach retract curves for noncontact situations: variation of the amplitude. The common parameters for the three curves are $HR=5\times10^{-27}$ J m, $u=0.9989$ $Q=470$, $k_c=40\ \mathrm{Nm}^{-1}$, resonance amplitude $A_0=16$ nm, working amplitude $A_{\mathrm{free}}=11.2$ nm. Arrows indicate the size of the loop hysteresis. The continuous line is calculated without any additional dissipation [Eq. (3)]. The two other curves are calculated by solving Eqs. (6a) and (6b) with a Mapple routine, with a local stiffness $k=1.2\ \mathrm{Nm}^{-1}$ (open circle) and $k=0.6\ \mathrm{Nm}^{-1}$ (filled circle). Note that when the dissipation increases, the cycle of hysteresis reduces and almost disappears for the highest dissipation. Also, the amplitude jump during the approach is reduced by half the nanometer. (b) Variations of the phase corresponding to the three cases displayed in (a). For the two first cases, at the bifurcation spot the phase jumps below the $-\pi/2$ value, while for the highest dissipation the phase jumps to a value around $-70°$. The additional dissipation strongly reduces the nonlinear behavior of the oscillator and, in turn, the distortion of the resonance peak.

$$\sin\phi=-au, \tag{2b}$$

where $\phi$ is the phase of the oscillator, $A_0$ the resonant amplitude, and $a$, $d$, and $u$ are the reduced values with $a=A/A_0$, $d=D/A_0$, and $u=\omega/\omega_0$. $\kappa_a=HR/k_cA_0^3$ is a dimensionless parameter with $k_c$ the cantilever stiffness. Varying $\kappa_a$ with $A_0$ is equivalent to varying the strength of the attractive interaction: for example, a large (small) $A_0$ corresponds to a small (large) $\kappa_a$. Qualitatively the influence of the oscillation amplitude can be described as follows: for a given closest distance $D-A$, because the oscillation period is a constant, the time during which the tip is at proximity of the surface depends on the oscillation amplitude. At large oscillation amplitudes this time, which can be called a residence time, is smaller than the one at small oscillation amplitudes. Therefore, the average attractive interaction during an oscillation period is a function of the oscillation amplitude. With a first-order expansion, one can demonstrate that the average zero frequency force component varies as $1/\sqrt{A}$.[29]

Phase variations as a function of $D$ or $A$ are readily obtained with one of the two equations, while the use of the trigonometric relation $\sin^2+\cos^2=1$ gives the relationship between the distance $D$ and the oscillation amplitude $A$:

$$d_{A\pm}=\sqrt{a^2+\left(\frac{Q\kappa_a}{3\left(Q(1-u^2)\mp\sqrt{\frac{1}{a^2}-u^2}\right)}\right)^{2/3}}. \tag{3}$$

The signs plus and minus correspond to the two branches of a bistable state.[7,9,14,19] Consequently, at a given tip surface distance a bifurcation from one stable oscillation state to a bistable one occurs leading to jumps of amplitude and phase. Equation (3) for the amplitude and Eqs. (2a) or (2b) for the phase gives suitable expressions to describe qualitatively the experimental features. But, as noted in Ref. 17, Eqs. (1) and (2) are unable to reproduce quantitatively variations of the amplitude and the observed relationship between amplitude and phase jumps. Everything happens as if an additional dissipation was not considered when the tip approaches the surface.

In the Appendix is given the main mathematical development leading to the expression of an additional effective damping coefficient $\beta_{\mathrm{eff}}(\Delta,A)$, with $\Delta\propto D-A$, when the tip does not touch the sample. Strictly, the approach is only valid when the amplitude $A$ is kept constant and for $\Delta\ll A$. If $A$ varies, as it happens with the Tapping mode, the situation is more complicated. For example, for a soft material, one has to solve a self-consistent equation to take into account the amplitude evolution.[29] Nevertheless, if the condition $\Delta\ll A$ remains verified throughout the variation of the amplitude $A$, the approach given below might be of some use. This approximation can only be supported by the ability of the expressions (A6a) and (A6b) to describe the general behavior over a wide range of experimental conditions (Sec. III). Using the effective damping coefficient [Eq. (A6b)]:

$$\beta_{\text{eff}}(\Delta) \approx \left( \frac{\omega_0}{\pi k_c} \frac{(HR)^2}{36k} \frac{1}{\Delta^4} \frac{1}{A^2} \right), \tag{4}$$

then $\beta_{\text{eff}}(\Delta,A)$ is added in the Lagrangian:

$$\begin{aligned} L &= T - U + W \\ &= \frac{1}{2} m \dot{x}^2 - \left( \frac{1}{2} k x^2 - x f \cos(\omega t) - \frac{HR}{6(D-x)} \right) \\ &\quad - (\beta_0 + \beta_{\text{eff}}(\Delta,\overline{A})) \dot{x} x, \end{aligned} \tag{5}$$

where $\beta_0$ is the oscillator's damping coefficient when $D \to \infty$. $\overline{A}$ means that to calculate the effect of an additional dissipation, the oscillation amplitude included in the expression of $\beta_{\text{eff}}(\Delta,\overline{A})$, is not varied when the action is minimized and is a solution of the stationary state. The above approach is reminiscent of the one done to describe what is called a structural dissipation. Such dissipation can occur for the large deformation of a plate or a rod. In that case because the dissipation becomes a function of the deformation, this nonlinear behavior, which is different than the one considered with a Van der Pol oscillator, is often solved through the introduction of an effective damping coefficient in which is included a fixed deformation.

Applying the variational principle as described in Ref. 19 gives the set of two equations:

$$\cos\phi = Qa(1-u^2) - \frac{Q\kappa_a}{3} \frac{a}{(d^2-a^2)^{3/2}}, \tag{6a}$$

$$\sin\phi = -au\left( 1 + \frac{\beta_{\text{eff}}(\Delta,\overline{A})}{\beta_0} \right). \tag{6b}$$

Inserting the expression of $\beta_{\text{eff}}(\Delta,\overline{A})$ [Eq. (4)] in Eq. (6b) and combining Eqs. (6a) and (6b) to get the amplitude variation leads to a complex polynomial equation that can only be solved numerically. Therefore, uniquely Eq. (6b) is of some use to fit the experimental data, in the present case the phase variation, for noncontact situations.

In Fig. 1(a) is reported the variation of the amplitude with and without additional dissipation. Variation of the amplitude without dissipation is straightforwardly obtained with Eq. (3). Since there is no analytical expression available for the relationship between $D$ and $A$, theoretical curves including the noncontact dissipation have been solved numerically with Mapple using Eqs. (6a) and (6b). In accordance with the known result that dissipation reduces the influence of the nonlinear terms,[7] the most obvious effect is the drastic reduction of the size of the hysteresis loop as a function of the dissipation. The main difference appears at the bifurcation spot where the amplitude jump is slightly frustrated. In Fig. 1(b) are shown the calculated phases with the equation (2a) [or (2b)] and the equation (6b). As expected, a marked change on the phase behavior is shown when the effective damping coefficient due to the attractive interaction is included. The additional dissipation reduces the influence of the strength of the attractive interaction between the tip and the sample, thus the distortion of the resonance peak, so that the phase shift is significantly reduced below $-\pi/2$. For large dissipation, the phase jump disappears and the phase rotates continuously over the $-\pi/2$ value.

### B. Repulsive regime

The intermittent contact situation includes both attractive and repulsive interactions. Thus, a more complex situation occurs requiring an additional hypothesis.[19] Here it is assumed that the tip experiences a repulsive interaction during a short time of its oscillating period and the attractive interaction is averaged on the whole oscillating period. The assumption is only valid for small indentations, $A - D \ll A$. Practically, such an assumption corresponds to approach retract curves for which the slope giving the rate of variation of the amplitude versus the cantilever surface distance is equal to one.[19] The repulsive interaction is assumed to have a simple harmonic form $F_{\text{Repulsive}}[x(t)] = k_s[x(t) - D]$ with $k_s$ the contact stiffness. For small indentations, $(a-d) \ll a$, the calculation gives the couple of equations:[19]

$$\begin{aligned} \cos(\phi) &= Qa(1-u^2) + \frac{4\sqrt{2}}{3\pi} Q\kappa_s a \left( 1 - \frac{d}{a} \right)^{3/2} \\ &\qquad - \frac{Q\kappa_a}{6\sqrt{2}\tilde{d}_c^{3/2}\sqrt{a}}, \\ \sin(\phi) &= -ua, \end{aligned} \tag{7}$$

leading to the relationship between $d$ and $a$:

$$d_{AR} = a\left[ 1 - \left( \frac{3\pi}{4\sqrt{2}} \frac{Qa(u^2-1) + \sqrt{1-(ua)^2} + \dfrac{Q\kappa_a}{6\sqrt{2}\tilde{d}_c^{3/2}} \dfrac{1}{\sqrt{a}}}{Q\kappa_s a} \right)^{2/3} \right], \tag{8}$$

where the attractive contribution is evaluated for $d = a + \tilde{d}_c$, i.e., at the closest NC distance from the surface, where $\tilde{d}_c$ is the reduced coordinate of $d_c$, the contact distance between most of the organic materials, $d_c = 0.165$ nm:[30] $\tilde{d}_c = d_c/A_0$. The repulsive term contains the parameter $\kappa_s$, which is a reduced stiffness given by the ratio between the contact stiffness $k_s$ and the cantilever one $k_c$: $\kappa_s = k_s/k_c$.

As for noncontact situations, Eq. (7b) can be replaced by $\sin(\phi) = -au[1 + \beta_{\text{eff}}(\Delta,A)/\beta_0]$. For hard surfaces, the indentation depth into the surface becomes negligible, of the order of $d_c$, and the expression of $\beta_{\text{eff}}(\Delta,A)$ can be simplified. It is enough to consider a constant effective tip surface distance $\overline{\Delta}$ below which the action of the tip becomes significant, typically $\overline{\Delta} \approx 0.5$ nm. Doing so, it uniquely remains in the $\beta_{\text{eff}}$ expression an explicit dependence as a function of the variation of the amplitude:

$$\beta_{\rm eff}(A) \approx \left( \frac{\omega_0}{\pi k_c} \frac{(HR)^2}{36k} \frac{1}{\Delta^4} \frac{1}{A^2} \right). \tag{9}$$

Therefore the solution giving the relationship between $D$ and $A$ can be obtained by replacing in Eqs. (7) and (8) the product $au$ by $au[1+\beta_{\rm eff}(A)/\beta_0]$:

$$d_{AR} = a\left[ 1 - \left( \frac{3\pi}{4\sqrt{2}} \frac{Qa(u^2-1) + \sqrt{1 - \left[ ua\left( 1 + \frac{\beta_{\rm eff}}{\beta_0} \right) \right]^2} + \frac{Q\kappa_a}{6\sqrt{2}\tilde{d}_c^{3/2}} \frac{1}{\sqrt{a}}}{Q\kappa_s a} \right)^{2/3} \right], \tag{10a}$$

$$\sin(\phi) = -au\left( 1 + \left( \frac{Q}{\pi k_c} \frac{(HR)^2}{36k} \frac{1}{\Delta^4} \frac{1}{A^2} \right) \right), \tag{10b}$$

with the substitution $Q=\omega_0/\beta_0$. Equations (10) give the set of equations that should be able to describe more appropriately the experimental results in intermittent contact situations, thus explaining the discrepancy between the variation of amplitude and phase.

In Fig. 2 are reported the theoretical curves deduced from Eqs. (7) and (8) and from Eqs. (10). As soon as the slope is equal to one, that is, for $Q\kappa_s \gg 10$,[19] the variations of the amplitude become insensitive to the relative strength of the attractive interaction or the amount of additional dissipation. This is a direct consequence of the fact that a slope equal to one cannot discriminate between an infinite hard surface or a surface with a finite value of the local stiffness with $Q\kappa_s \gg 10$. In other words, amplitude curves are useful to extract quantitative values of the contact stiffness only when materials are soft enough to produce slopes smaller than one. Therefore, most of the information is obtained on the phase variations.

## III. COMPARISON WITH EXPERIMENTAL DATA

The resonance frequency is $\nu_0 = 185\,500$ Hz, the quality factor is $Q=470$, the experiments were performed at $u=\nu/\nu_0=0.9989$, corresponding to a phase $\phi=-45°$, and $A_{\rm free}=A_0/\sqrt{2}$ for $D\rightarrow\infty$, where $A_0$ is the amplitude at the resonance frequency. The AFM is set in a glove box in which the PPM of water is achieved allowing the OTCL to keep a stable behavior. Two surfaces have been investigated: a silica surface and a silica grafted with aminopropylsilane (APTES). The surface treatment is given in detail elsewhere,[31] APTES was chosen because of its ability to stick DNA molecules onto a surface, while the silica surface is used as a reference. The experiments were performed with the same tip without any evidence of change of the size of the tip or change of the tip pollution. Since the strength of the attractive interaction is governed by the oscillation amplitude, approach retract curves were recorded at different working amplitudes $A_{\rm free}$, ranging from 53 down to 4 nm.

### A. Noncontact situations: Evaluation of the attractive interaction between the tip and the surface

The first step is to estimate the product $HR$ for the two surfaces. A qualitative picture is readily obtained by looking at the amplitude at which the phase crosses the $-\pi/2$ value (Fig. 3). The phase variation at the bifurcation spot can be understood as follows: when the oscillator experiences a dominant attractive regime, the resonance peak distorts toward the low frequency.[7,9,14] Because we use a drive frequency slightly below the resonance one, the amplitude jump occurs by crossing the $-\pi/2$ value. When the oscillator experiences a dominant repulsive regime, the resonance peak distorts mainly toward the high frequency and the phase remains above the $-\pi/2$ value.

The strength of the attractive interaction, the dimensionless parameter $\kappa_a$ scaling as $HR/A_0^3$, will decide whether or not the oscillator is in a dominant repulsive or attractive regime. Therefore, at low oscillation amplitudes one may expect to have a large $\kappa_a$ thus an attractive regime, while at large oscillation amplitudes, $\kappa_a$ may become small enough so that a dominant repulsive regime controls the behavior of the oscillator. Consequently, for two different surfaces and with the same oscillation amplitude, if one surface indicates a dominant attractive regime while the other indicates a dominant repulsive one, that will mean that the former surface does have a larger product $HR$ than the latter one. In addition, if we do use the same tip, thus the same radius $R$, such a comparison provides direct information about the relative strength of the Hamaker constant of the two surfaces characterizing the tip sample interaction.

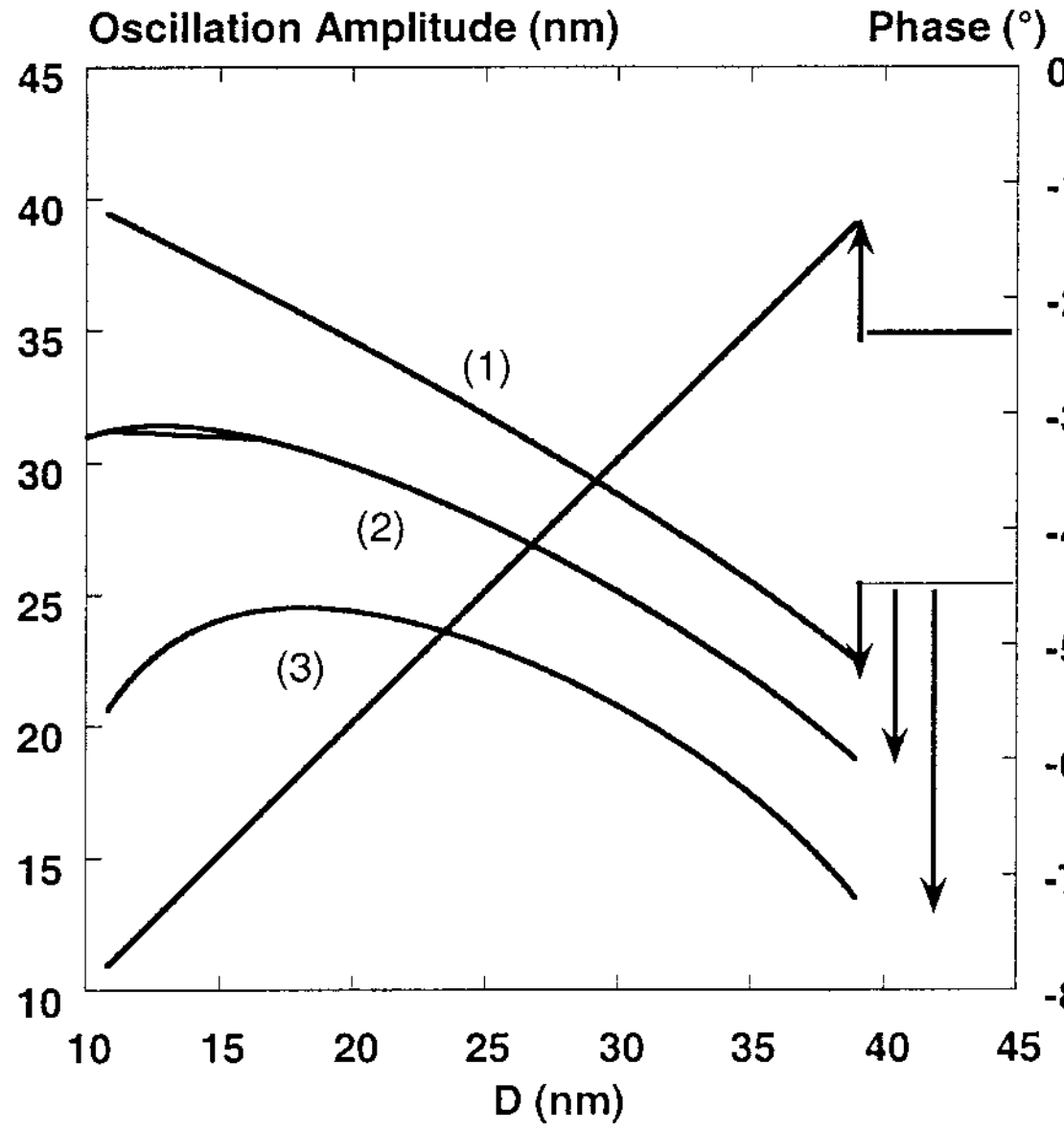

FIG. 2. Amplitude and phase curves computed with Eqs. (10) corresponding to intermittent contact situations. The parameters are identical to the ones used for Figs. 1, except the resonance amplitude $A_0=50$ nm and the working amplitude $A_{\rm free}=35$ nm. The phase curve (1) is calculated with Eq. (7b) (without additional dissipation) while the phase curves (2) and (3) are calculated with Eq. (10b). Contrary to the noncontact situations, the phase jump increases as the additional dissipation increases.

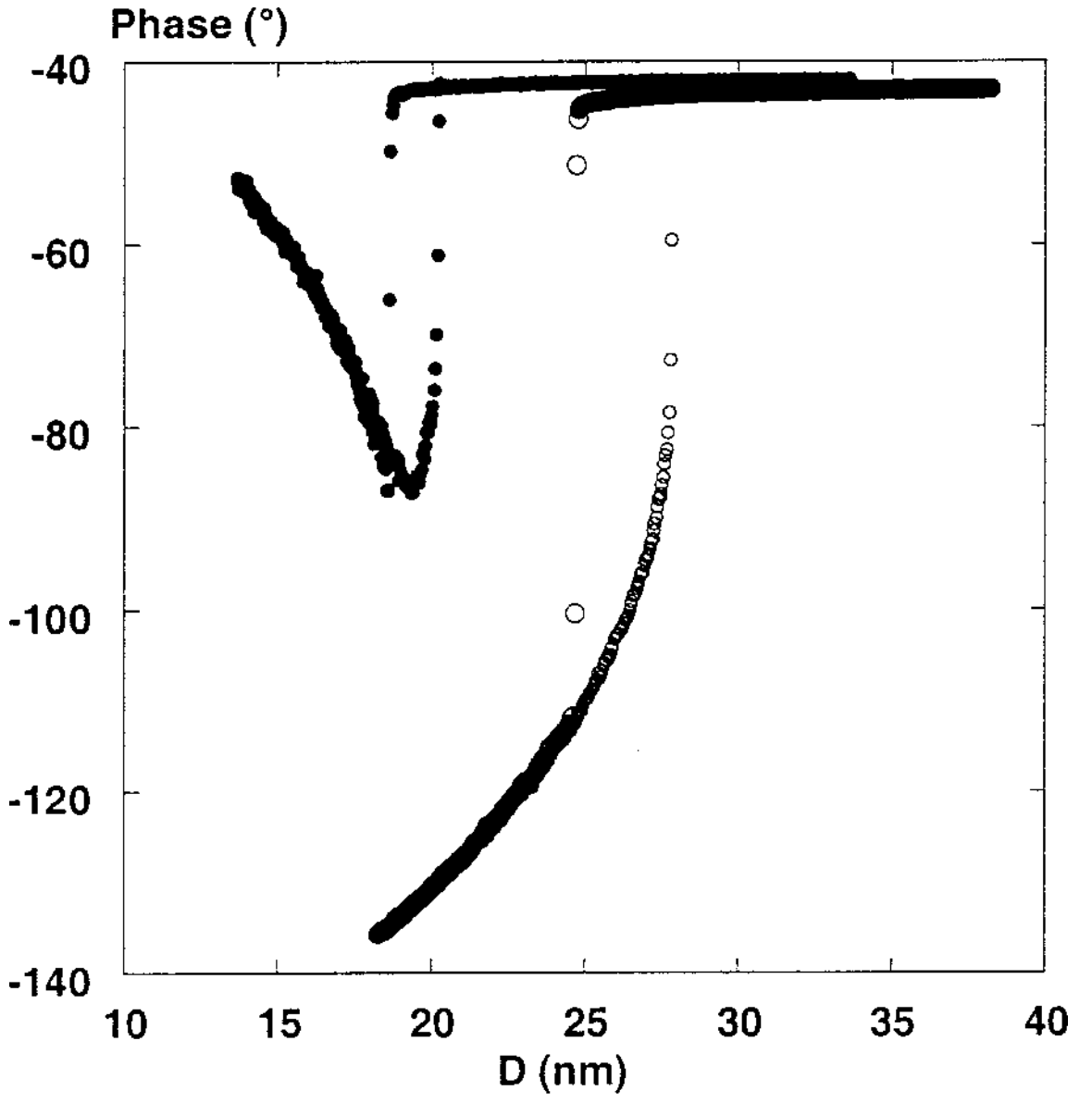

FIG. 3. Phase experimental curves obtained on silica (filled circle) with a working amplitude $A_{\mathrm{free}}=15$ nm and APTES surface (open circle) with a working amplitude $A_{\mathrm{free}}=22$. The variation of the phase indicates a noncontact situation for the APTES and an intermittent contact situation for the silica surface. In spite of the fact that the working amplitude is larger for the APTES than that of the silica, the tip surface interaction is large enough to reduce the amplitude at the proximity of the surface without the need of a contact.

The amplitudes at which the NC situation occurs are $A_{\mathrm{free}}=32$ nm ($A_0=44$ nm) for the grafted surface with the amine group and $A_{\mathrm{free}}=13$ nm ($A_0=19$ nm) for the silica. Because the tip's size is a constant, this result immediately indicates that the interaction is much larger for the grafted surface, particularly for silanes with amine groups in which an additional Debye interaction due to the amonium group is present.

An attempt to obtain a more quantitative evaluation can be done by comparing the experimental curves to the theoretical ones. As shown in Fig. 1(a) the energy loss due to the attractive interaction strongly modify the hysteresis loop and, to a lesser extent, acts on the variation of the amplitude near the surface. However, the very beginning of the amplitude variation, corresponding to the increase of the amplitude before the bifurcation spot, is only slightly modified. Therefore, the use of Eq. (3) becomes of some help in evaluating the product $HR$. We focus on curves for which the noncontact situation occurs for the whole variation of the amplitude and phase. In Figs. 4 are reported several comparisons between experimental curves and theoretical ones. With a cantilever stiffness $k_c=40\,\mathrm{Nm}^{-1}$, the products $HR$ are $5\times10^{-27}$ and $11.5\times10^{-27}$ J m for the silica and the grafted surface, respectively. The estimated error is difficult to evaluate; one may also get a rather good agreement with values 20% higher.

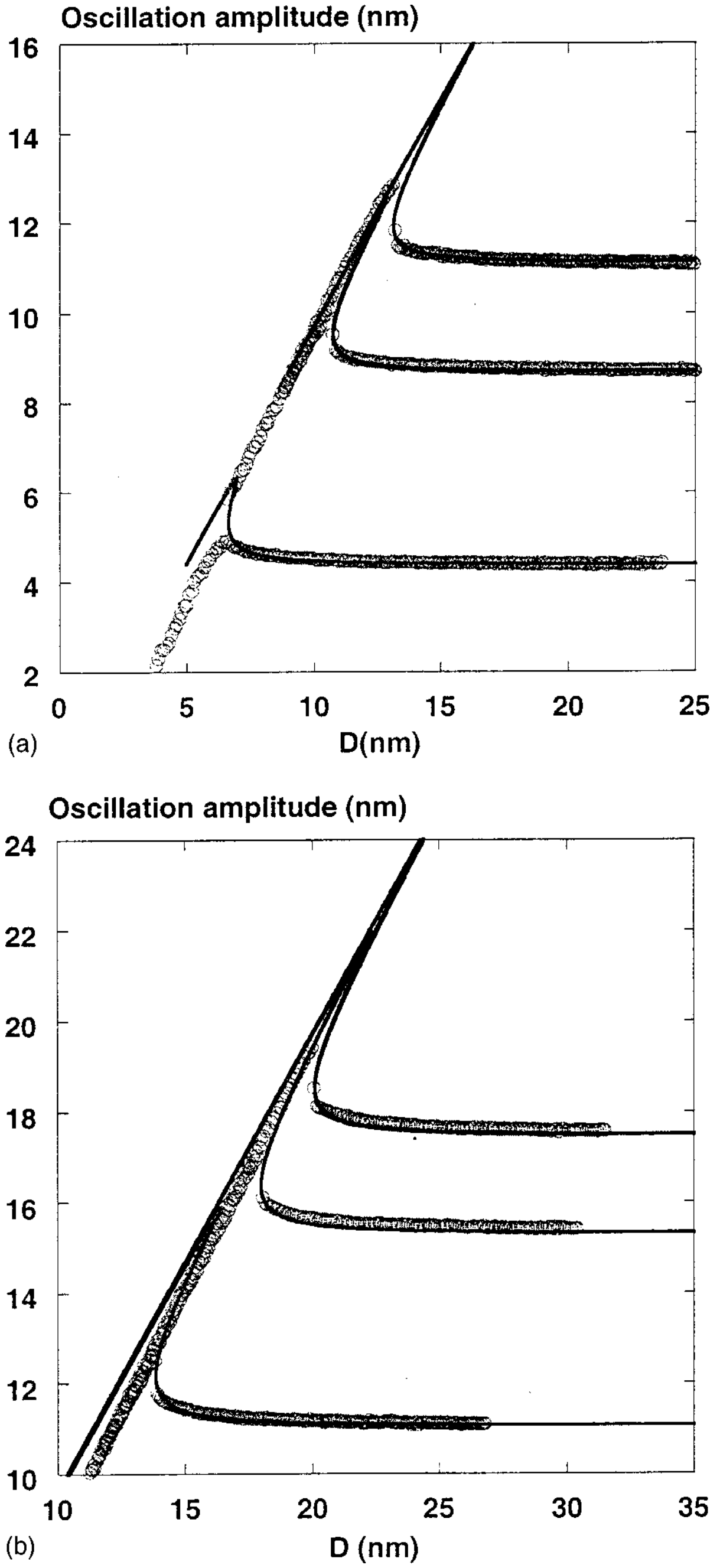

FIG. 4. A comparison between experimental approach curves and theoretical ones calculated without including an additional dissipation [Eq. (3)]. (a) Silica surface with $A_{\mathrm{free}}=4$ nm, $A_{\mathrm{free}}=9$ nm, $A_{\mathrm{free}}=11$ nm. (b) APTES surface with $A_{\mathrm{free}}=11$ nm, $A_{\mathrm{free}}=16$ nm, $A_{\mathrm{free}}=18$ nm. The theoretical curves are calculated with the experimental parameters $Q$, $u$, and $A_0$ (see the text) and $k_c=40\,\mathrm{Nm}^{-1}$, $HR=11.5\times10^{-27}$ J m for the APTES and $HR=5\times10^{-27}$ J m for the silica.

The fit with Eq. (3) also provides the opportunity to locate the surface. The example shown in Fig. 5 gives a bifurcation spot at 1.7 nm while after the bifurcation the closest distance is about 1.3 nm (Fig. 5). These approach retract curves correspond to a large attractive interaction on the grafted surface with a working amplitude $A_{\mathrm{free}}=9$ nm. The comparison with a theoretical curve including the dissipation does show a noticeable difference. The experimental curve exhibits a slope larger than the theoretical one. Such a situation occurs if the sample displacement under the action of the tip becomes large enough to modify significantly the tip sample distance. Note that the local stiffness of a surface is the product of an intrinsic property, the elastic modulus $G$, and the diameter $\phi$ of the area involved in the interaction, thus $k\approx G\phi$. This is a general problem of the local probe method, which is sensitive to intrinsic properties of materials with a number of elementary units difficult to evaluate. In the present calculation, the value used for the fit, $k=1\ \mathrm{Nm}^{-1}$, might correspond to an elastic modulus of $10^8\ \mathrm{Nm}^{-2}$ if $\phi=10$ nm or $10^9\ \mathrm{Nm}^{-2}$ if $\phi=1$ nm.

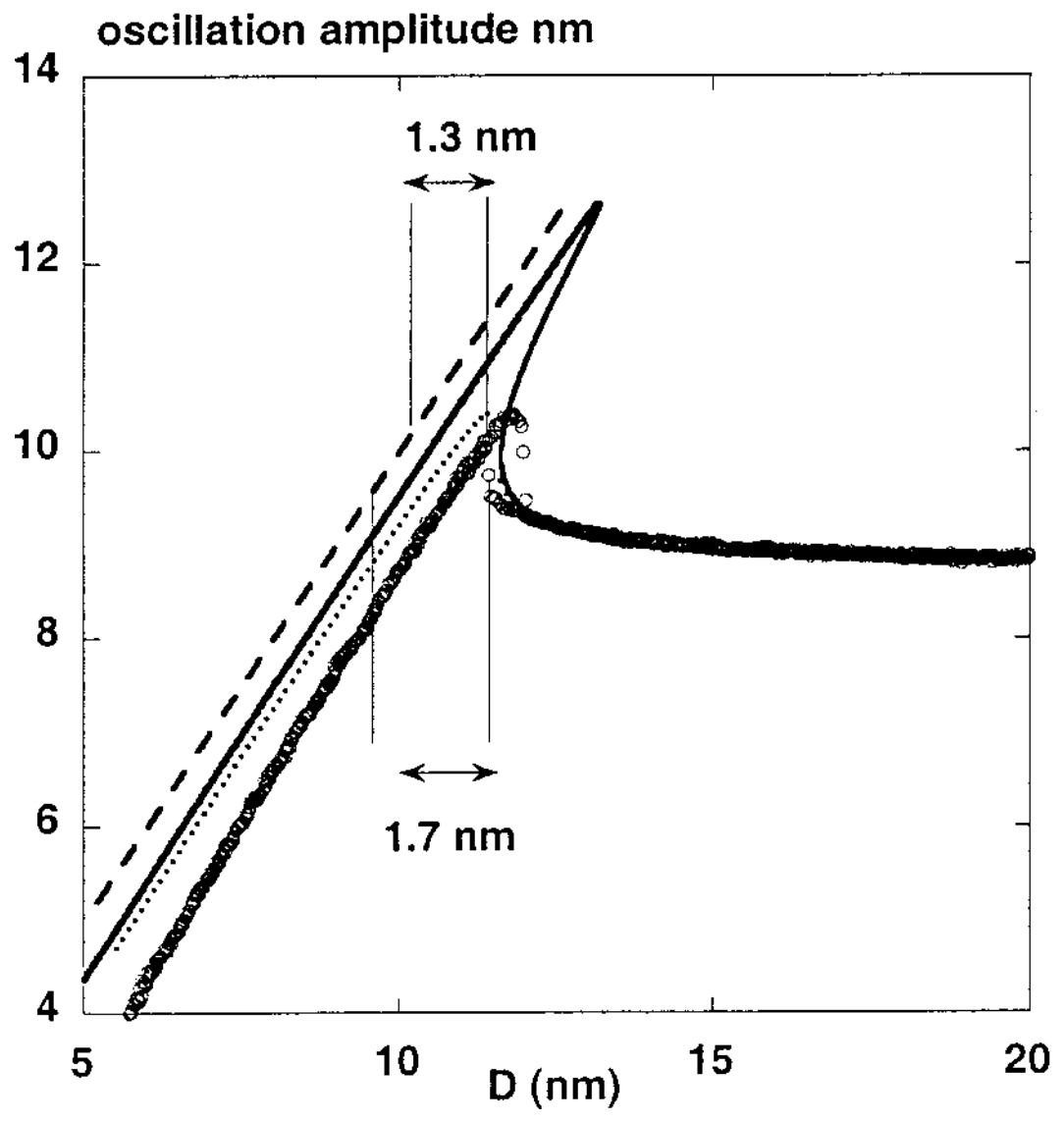

FIG. 5. A comparison between the experimental amplitude variation of the APTES (open circle) at the working amplitude $A_{\mathrm{free}}=9$ nm and theoretical curves. Curves without dissipation calculated with Eq. (3) and the experimental parameters $u=0.9989$, $Q=470$, $k_c=40\ \mathrm{Nm}^{-1}$, resonance amplitude $A_0=13$ nm, working amplitude $A_{\mathrm{free}}=9$ nm and the input parameter $HR=11.5\times 10^{-27}$ J m (continuous line). The curve including the dissipation with $k=1\ \mathrm{Nm}^{-1}$ (dotted line). The location of the surface (dashed line).

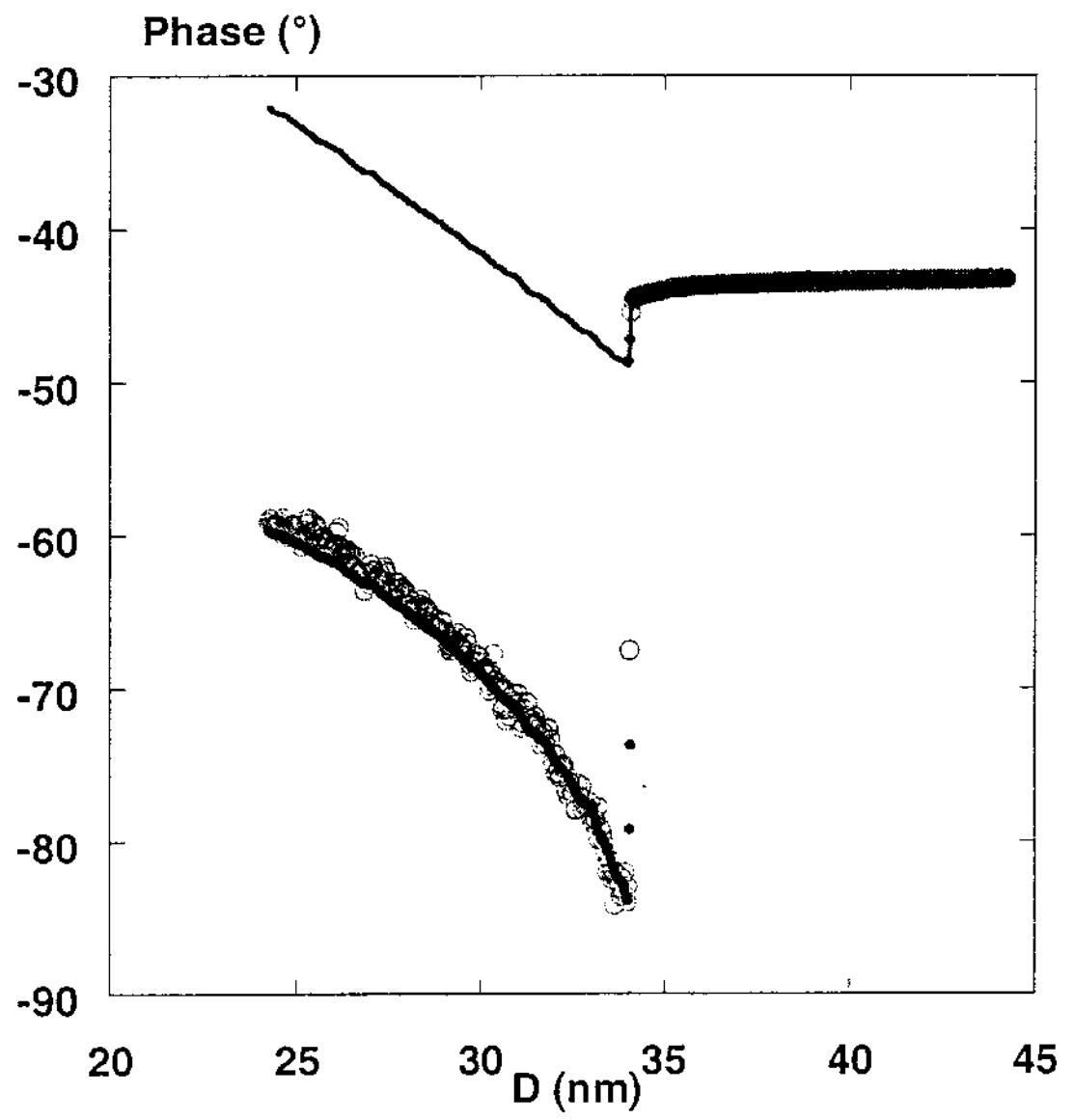

FIG. 6. Example showing the influence of the noncontact dissipation for the intermittent contact situation. The continuous line is obtained by calculating the phase with $\phi=\sin^{-1}(uA/A_0)$ [Eq. (7b)], the filled circles are given by a fit with Eq. (10b). The experimental data (open circle) correspond to the APTES surface with $A_{\mathrm{free}}=31$ nm.

As stated in Sec. II, including the surface displacement, does not lead to a simple analytical expression to describe the dissipation. While the surface displacement under the action of the tip is the driving term controlling the amount of additional dissipation, the theoretical description neglects the magnitude of this surface displacement. Such an assumption may become a rough one when a strong attractive interaction occurs as it happens at a low oscillation amplitude. As it is shown below, for a weaker attractive interaction leading to intermittent contact situations, the analytical expression used to fit the experimental data gives a good agreement.

Equation (6b) contains three unknown parameters: the product $HR$, the tip–surface distance $\Delta$, and the surface mechanical response $k$. The first parameter is now evaluated, the second is also approximately estimated, and the third is determined with the intermittent contact situations by setting an arbitrary value of $\Delta$ (see Sec. III B). The dominant repulsive regime, with well-defined intermittent contact situations, is easier to fit because of a $D-A$ distance remaining constant throughout the variation of the amplitude. With the dominant attractive regime, as mentioned above, one has to take into account a possible contribution of the elastic displacement of the surface, but also a contribution due to a slight contact with the surface, which is not taken into account. Nevertheless, while at intermediary amplitudes the agreement is fairly good, the overall behavior is quite well reproduced.

## B. Intermittent contact situations

The action of the oscillating nanotip is described as a pulse based on time scale considerations, with the basic assumption of a rectangular periodic function sustained by the fact that the force $HR/6[D-x(t)]^2$ can be suitably replaced by $HR/6\Delta^2\cdot\Delta$ is a fixed effective distance between the tip and the surface giving an order of magnitude of the strength of the attractive interaction. To simplify our evaluation, we consider an effective distance $\Delta=0.5$ nm. Also, because the slope on those surfaces is equal to one, the indentation depth is very small, and we set arbitrarily the contact distance at $D-A=0.165$ nm at the jump value of the amplitude during the approach (such a procedure gives a good estimation of the surface location for a hard surface). Therefore, for intermittent contact situations, $D-A$ is less than the percent of the amplitude, and $\Delta\ll A$. Using the results obtained in Sec. III A and with $\Delta=0.5$ nm gives an average attractive interaction of $3.2\times 10^{-9}$ and $7.6\times 10^{-9}$ N for the silica and grafted surface, respectively.

In Fig. 6 is displayed an observed phase variation and the calculated curves using Eq. (7b) and Eq. (10b). The difference is striking; the influence of the additional dissipation due to the attractive interaction is unambiguously shown. Two main effects are emphasized: the first one concerns the phase jump, Eq. (7b) gives a jump of 5° while the experimental results and Eq. (10b), with the adjusted mechanical response $k$, gives a jump of 40°. The second one concerns the variation of the phase. Equation (10b) reproduces with a good agreement the phase variation. This good correspondence suggests that the influence of the decrease of the oscillation be correctly ascribed through the power law $A^{-2}$.

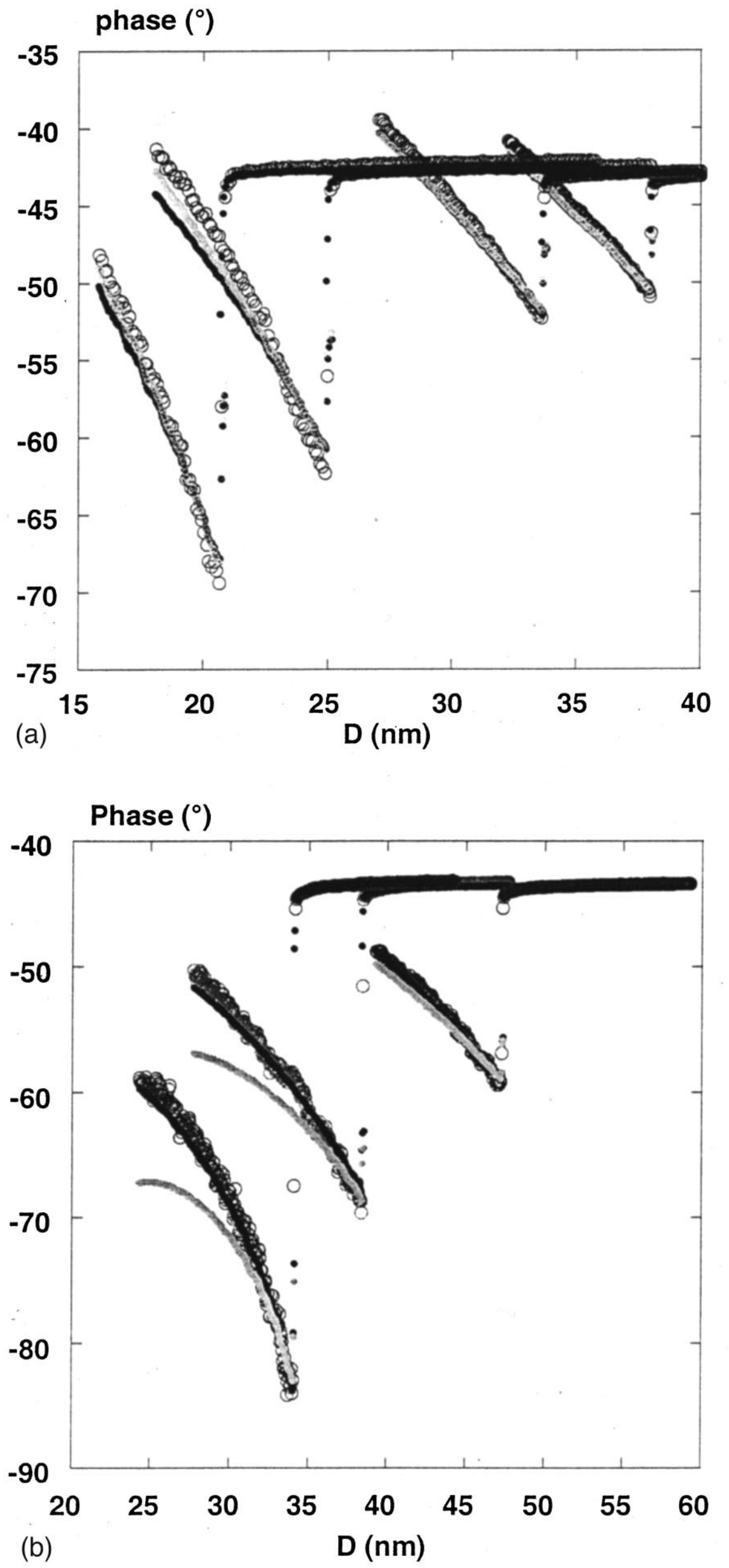

FIG. 7. Variations of the phase in intermittent contact situations for different working amplitudes. A comparison with calculated curves: Black dots power law $A^{-2}$, Eq. (10b); gray dots power law $A^{-5/2}$ [Eq. (7a)]. Silica surface, $A_{\mathrm{free}}$: 35, 31, 22, 18 nm (Fig. 8a), APTES surface $A_{\mathrm{free}}$: 44, 35, 31 nm (Fig. 8b).

A comparison between calculated curves and experimental ones are displayed in Fig. 7. Also are included curves obtained with the power law $A^{-5/2}$. Fits performed on the curves measured on the silica surface do not allow the two power laws to be separated unambiguously [Fig. 7(a)], while the comparison done with the curves recorded on the grafted surface provides an unambiguous answer [Figs. 7(b)]. This may be due to a much larger strength of the attractive interaction on the grafted silica surface, thus providing the opportunity to discriminate between the two regimes. The differ-ence between the two power laws is due to the ratio between the characteristic time of the surface and the residence time $\tau_{\mathrm{res}}$ (see the Appendix). An $A^{-5/2}$ power law will be observed for rather slow relaxation processes.

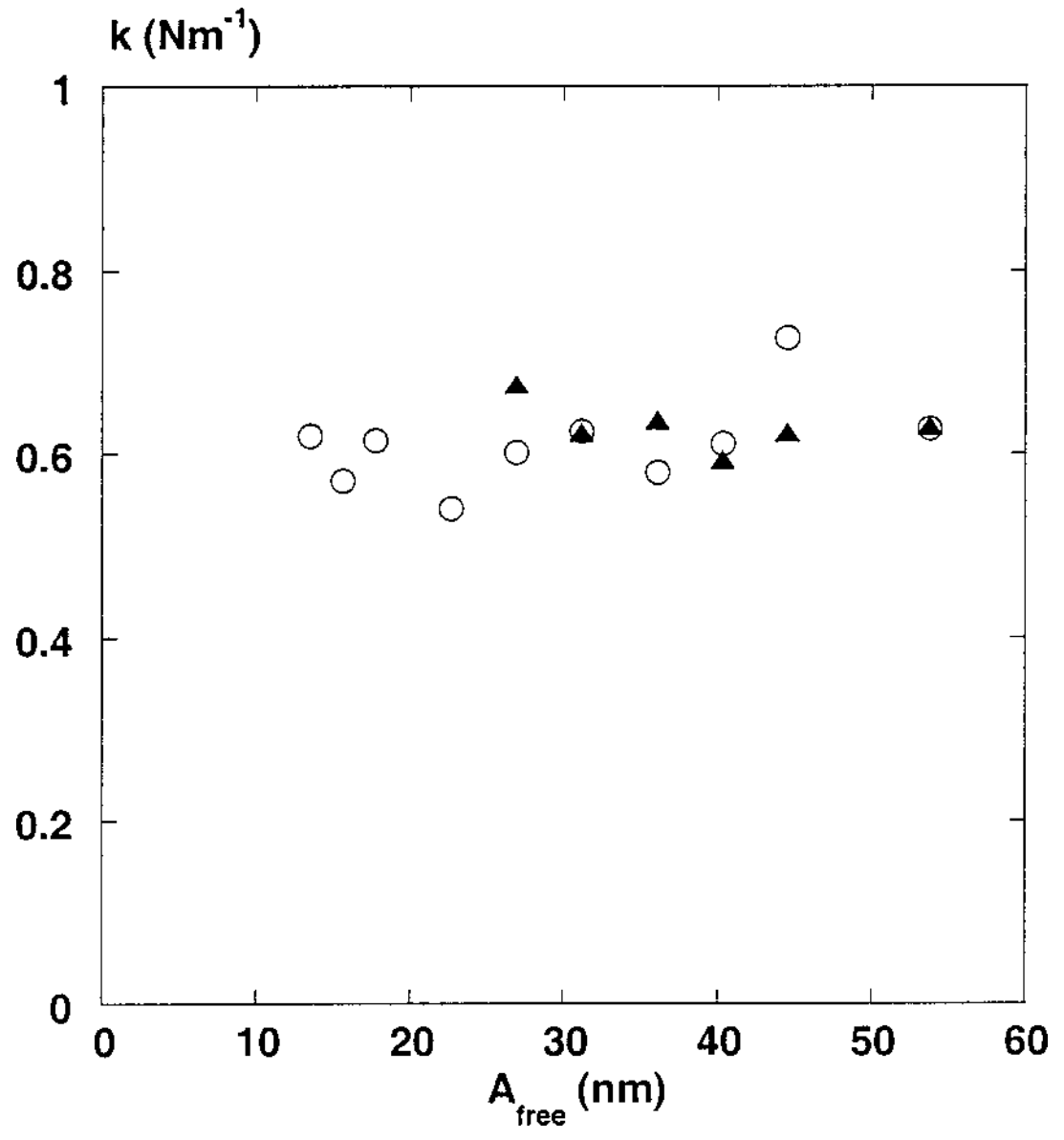

FIG. 8. Variation of the mechanical susceptibility $k$ of the surface controlling the amount of noncontact dissipation. The values are obtained from fits with Eq. (10b) (see Fig. 7) with $\Delta=0.5$ nm. Open circle: silica surface; filled triangle: grafted surface.

The mechanical susceptibilities extracted from the fits exhibit a plateau throughout the range of amplitude investigated (Fig. 8). This is in good agreement with the prediction of the model using an average effective distance $\Delta$ constant and a negligible contribution of the elastic displacement of the surface. This result strongly supports the working hypothesis employed to derive the equation (10b). Also, it does appear that the large increase of dissipation on APTES is mainly due to an increase of the strength of the attractive interaction through the square of the product $HR$ [Eq. (9)]. Therefore, the same kind of mechanical susceptibilities control the amount of additional dissipation for the two surfaces. Such a result is not really surprising since both ends of the short APTES molecules can interact with the silica surface, thus giving a surface morphology nearly identical with tightly bounded molecules. In addition, it is known that two or three water layers in a glassy or ''solid'' state are strongly adsorbed onto the silica surface,[32] thus modifying the mechanical properties.[33] One can expect that these few layers with an amorphouslike behavior might be the origin of the dissipating effect in the dominant attractive regime.

While the above analytical expressions are useful to compare the properties of different surfaces, there remain some difficulties related to the use of the local probe method. One is that the parameters fitted are always the product of two quantities. For an estimation of the strength of the attractive interaction, the product $HR$ is evaluated, such that only a guess of the tip size gives an estimation of the Hamaker constant. For example, a radius of the tip of 50 nm leads to a Hamaker constant of $10^{-19}$ J. In the same way, the fit of the additional dissipation gives a value of the product $\Delta^4 k$. Thus, the choice of an average distance of 0.5 nm gives a mechanical susceptibility of 0.6 $\mathrm{Nm}^{-1}$ while a value of 1 nm will give a value 16 times smaller.

## IV. CONCLUSION

The present work was an attempt to make a quantitative analysis of the variation of the phase and of the oscillation amplitude in the Tapping mode for hard surfaces. To do so, an additional dissipation due to the attractive interaction between the tip and the surface is included in the Lagrangian formalism. A simple model based on a pulse rectangular function to describe the action of the tip and a viscoelastic behavior to take into account the mechanical response of the surface is used. This simplified approach allows analytical solutions to be derived. In spite of these crude assumptions, this phenomenological approach is able to reproduce most of the observed features. Particularly, the expressions obtained are able to reproduce with a good agreement the relationship between phase and amplitude when the tip is at proximity of the surface or in intermittent contact situations. The ability to fit experimental variations of the oscillation amplitude and phase as a function of the tip surface distance should give us the opportunity to obtain more accurate information on the properties of the underneath surface.

## APPENDIX

The time during which the tip is close to the surface is called the residence time; following the approach given in Ref. 29, one can consider an average tip–sample distance $\Delta$ at proximity of the surface such that the residence time is given by $\tau_{\mathrm{res}} \approx (T/\pi)\sqrt{2\Delta/A}$ with the period $T = 2\pi/\omega_0$. Therefore, the action of the oscillating tip can be described as a rectangular periodic function of width $\tau_{\mathrm{res}}$ and height $F_{\mathrm{ext}} = HR/6\Delta^2$.[29] For fast relaxation times of the protuberance $\beta^{-1}$, with $\beta \gg \omega_0/2\pi$, the action of the oscillating tip can be described as a pulse of width $\tau_{\mathrm{res}}$ and one can use an integration instead of the Fourier series. The dissipated energy due to the attractive interaction between the tip and the sample is

$$\langle E_{\mathrm{diss}} \rangle_T = \int_0^\infty \omega \chi''(\omega) |f_\omega|^2 \frac{d\omega}{\pi}. \tag{A1}$$

With the Fourier coefficient $f_\omega = 2F_{\mathrm{ext}}(\sin[\omega(\tau_{\mathrm{res}}/2)]/\omega)$ at the frequency $\omega$ and $\chi''(\omega)$ the imaginary part of the generalized susceptibility $\chi = \chi' + i\chi''$. Equation (A1) expresses that part of the work performed by the oscillating tip on the sample is not restored to the oscillator and vanished in the bulk. Due to the attractive tip surface interaction, which can be as large as 1 nN, a surface displacement, the growth of a nanoprotuberance, occurs with a phase delay if the surface is not a pure elastic one. Such an approach remains correct if the vertical surface displacement remains small.[26] From Eq. (A1) one gets the result that the amount of dissipated energy varies as the square of the attractive force (see also Ref. 34); therefore a dependence in $1/\Delta^4$ is expected [see Eq. (A3)].

The next step is to express $\chi''(\omega)$. To do so we use a simple phenomenological model describing the surface properties with a viscoelastic mechanical response that can be represented with a spring constant $k$ ($\mathrm{Nm}^{-1}$) and a damping factor $\gamma$ ($\mathrm{kg\,s}^{-1}$) in parallel. For example, we may consider that the locality of the coupling between the oscillating tip and the surface is described by a local elastic response of the surface with a stiffness $k$ that is coupled to a surrounding medium of mass $M$ with an intrinsic molecular relaxation time $\tau_m = \beta^{-1}$, thus a damping term $\gamma = M\beta$. With this simple description of the surface, a highly dissipating material with large relaxation times has a surface displacement proportional to $1/\gamma$ while a weakly dissipating material with short relaxation times has a surface displacement proportional to $1/k$. Thus, the corresponding amount of dissipated energy, which is a function of the surface deformation, must show a similar behavior [see Eqs. (A6)] $\chi''(\omega)$ is given by

$$\chi''(\omega) = \frac{\omega\gamma}{k^2 + \gamma^2\omega^2}. \tag{A2}$$

Inserting (A2) in expression (A1) leads to the result

$$\langle E_{\mathrm{diss}} \rangle_T = \frac{(HR)^2}{36\Delta^4} \frac{1}{k} \left[ 1 - \exp\left( -\frac{\tau_{\mathrm{res}} k}{\gamma} \right) \right]. \tag{A3}$$

Two asymptotic regimes, which are determined by the values of the ratio $\tau_{\mathrm{res}} k/\gamma$, are extracted from Eq. (A3). The asymptotic regimes correspond to two limiting cases of the sample mechanical response.

For $\tau_{\mathrm{res}} k/\gamma \ll 1$, Eq. (A3) can be by replaced by

$$\langle E_{\mathrm{diss}} \rangle_T \approx \frac{(HR)^2}{36\Delta^4} \frac{\tau_{\mathrm{res}}}{\gamma} \approx \frac{(HR)^2}{18\Delta^{7/2}} \frac{1}{A^{1/2}} \frac{\sqrt{2}}{\gamma\omega_0}, \tag{A4a}$$

where $\tau_{\mathrm{res}}$ has been substituted by $(2/\omega_0)\sqrt{2\Delta/A}$,

while for $\tau_{\mathrm{res}} k/\gamma \gg 1$, one gets

$$\langle E_{\mathrm{diss}} \rangle_T \approx \frac{(HR)^2}{36\Delta^4} \frac{1}{k}. \tag{A4b}$$

Equation (4b) means that the average dissipation energy per pulse is mainly governed by the magnitude of the local stiffness, while Eq. (4a) exhibits explicitly the viscous process. Equation (4a) would be more suitable for material having a dominant friction behavior and, or dissipating processes due to diffusion motion, while Eq. (4b) is more likely to describe hard surface behavior with phonon assisted dissipation.

We now have to express that the oscillator loses this energy during a period. The simplest way to describe the loss of energy is to use an equivalent damping coefficient $\beta_{\mathrm{eq}}$ that becomes a function of the closest tip–sample distance $\Delta$. The energy dissipated during a period is given by

$$\langle E_{\mathrm{diss}} \rangle_T = m\pi\beta_{\mathrm{eq}}(\Delta)\omega_0 A^2 = k_c \pi \beta_{\mathrm{eq}}(\Delta) \frac{A^2}{\omega_0}, \tag{A5}$$

where $k_c$ is the cantilever stiffness. Combining Eqs. (A4) and (A5) gives an expression of the equivalent damping coefficient:

$$\beta_{\mathrm{eq}}(\Delta) \approx \left( \frac{\sqrt{2}}{\pi k_c} \frac{(HR)^2}{18\gamma} \frac{1}{\Delta^{7/2}} \frac{1}{A^{5/2}} \right), \tag{A6a}$$

$$\beta_{\mathrm{eq}}(\Delta)\approx\left(\frac{\omega_0}{\pi k_c}\,\frac{(HR)^2}{36k}\,\frac{1}{\Delta^4}\,\frac{1}{A^2}\right). \tag{A6b}$$